\documentclass[10pt,journal,twoside,twocolumn]{IEEEtran}
\usepackage{algorithm}
\usepackage{algorithmic}

\usepackage{color}

\usepackage{graphicx}
\graphicspath{{figures/},{pics/}}

\usepackage{amsmath,bm}  
\usepackage{amssymb} 

\usepackage{url}
\usepackage{cite} 

\title{Statistical CSI-Based  Transmission Design for Reconfigurable Intelligent Surface-aided Massive MIMO Systems with Hardware Impairments}

\author{
	Jianxin Dai, Feng Zhu, Cunhua Pan, Hong Ren and Kezhi Wang
	\thanks{(Corresponding author: Cunhua Pan).}
	\thanks{This work was supported by the Postgraduate Research \& Practice Innovation Program of Jiangsu Province, China (Grant No. SJCX21\_0268), and the National Key Research and Development Project under Grant 2019YFE0123600.
		
		J. Dai is with School of Science, Nanjing University of Posts and Telecommunications, Nanjing 210096, China.(email:daijx@njupt.edu.cn). 	
	F. Zhu is with School of Science, Nanjing University of Posts and Telecommunications, Nanjing 210096, China.	(email:1220014024@njupt.edu.cn).
	C. Pan is with the School of Electronic Engineering and Computer Science at Queen Mary University of London, London E1 4NS, U.K. (e-mail:k.zhi, c.pan@qmul.ac.uk).
	H. Ren is with the National Mobile Communications Research Laboratory, Southeast University, Nanjing 210096, China. (hren@seu.edu.cn).
	K. Wang is with Department of Computer and Information Sciences, Northumbria University, UK. (e-mail: kezhi.wang@northumbria.ac.uk).
			}
}

\begin{document}
	\maketitle	
	\markboth{}{}
	
	\begin{abstract}
	We consider a reconfigurable intelligent surface (RIS)-aided  massive multi-user multiple-input multiple-output (MIMO) communication system with \textcolor[rgb]{0,0,0}{transceiver hardware impairments (HWIs) and RIS phase noise}. Different from the existing contributions, the phase shifts of the RIS are designed based on the long-term angle informations.  Firstly, an approximate analytical expression of the uplink achievable rate is derived. Then, we use genetic algorithm (GA) to maximize the sum rate and the minimum date rate. Finally, \textcolor[rgb]{0,0,0}{we show that it is crucial to take HWIs into account  when designing the phase shift of RIS.}
	\end{abstract}
	

	\IEEEpeerreviewmaketitle

\vspace{-0.2cm}
	\section{Introduction}
	With the rapid development of wireless communication technology,  reconfigurable intelligent surface (RIS) has been recognized
	as a revolutionary technology for future wireless  communication
	systems \cite{MD2019Smart}.
	An RIS consists of passively reflecting elements and each element can independently induce certain phase shift changes to the incoming signal \cite{8796365},\cite{8910627}.
RIS-aided communication systems have been extensively studied  in \cite{2020Power,9355404}. The authors in \cite{2020Power} investigated an RIS-assisted massive \textcolor[rgb]{0,0,0}{multiple-input multiple-output (MIMO)} system, derived a closed-form expressions of the rate. In \cite{9355404}, the performance of RIS-assisted  MIMO systems with direct links was studied based on statistical channel state information (CSI). 

	
	Recently, the transceiver hardware impairments (HWIs) were considered in \cite{9239335,2021Intelligent,9322510,9390410}. Specifically, in \cite{9239335}, the authors focused on an RIS-assisted multi-antenna communication system with transceiver HWIs, and designed the transmit and reflecting beamforming. In \cite{2021Intelligent}, an RIS-aided  communication system  was studied based on the imperfect hardware, where the authors derived the spectral efficiency by considering imperfect CSI and presented a general methodology for the RIS's reflecting beamforming (RB) optimization. \textcolor[rgb]{0,0,0}{In \cite{9322510}, the authors derived the optimal receive combining and transmit beamforming vectors, and provided the analytical upper and lower bounds on the maximal energy efficiency. In \cite{9390410}, the authors derived the closed-form expression of the average achievable rate with HWIs.}
	
	However, the HWIs  under the scenario of RIS-aided massive MIMO have not been investigated. Against this background, \textcolor[rgb]{0,0,0}{we consider the uplink transmission of an RIS-aided massive MIMO system based on the statistical CSI, where transceiver has  additive HWIs and RIS has phase noise.} We consider the availability of statistical CSI because it changes more slowly than instantaneous CSI and it can also significantly relax  the necessity  of frequently reconfiguring the RISs \cite{124578},\cite{9090356}. \textcolor[rgb]{0,0,0}{Besides, by increasing the element spacing of RIS, we can avoid spatial correlation.} 
	 Specifically, an approximate analytical expression of the rate is derived. Then, to maximize the rate, GA is used to optimize the phase shifts.  \textcolor[rgb]{0,0,0}{Finally, we show that it is crucial to take HWIs into account  when designing the phase shift of RIS.}
	
%
	\textit{Notations}: $||\cdot||$ stands for the  $l_2$ norm of a vector.  $\textbf{x} \in \mathcal{CN}(\mathbf{a},\mathbf{\Sigma})$ denotes that $\textbf{x}$ is a complex Gaussian random vector with mean $\mathbf{a}$ and covariance matrix $\mathbf{\Sigma}$.
	$\widetilde{diag}\{\cdot\}$ denotes a diagonal matrix whose diagonal elements are the same as the original matrix.

	%
	\vspace{0.2cm}
	\section{System Model} 
			 \vspace{-0.5cm}
	\begin{figure}[h] 
		 \centering{\includegraphics[scale=0.4]{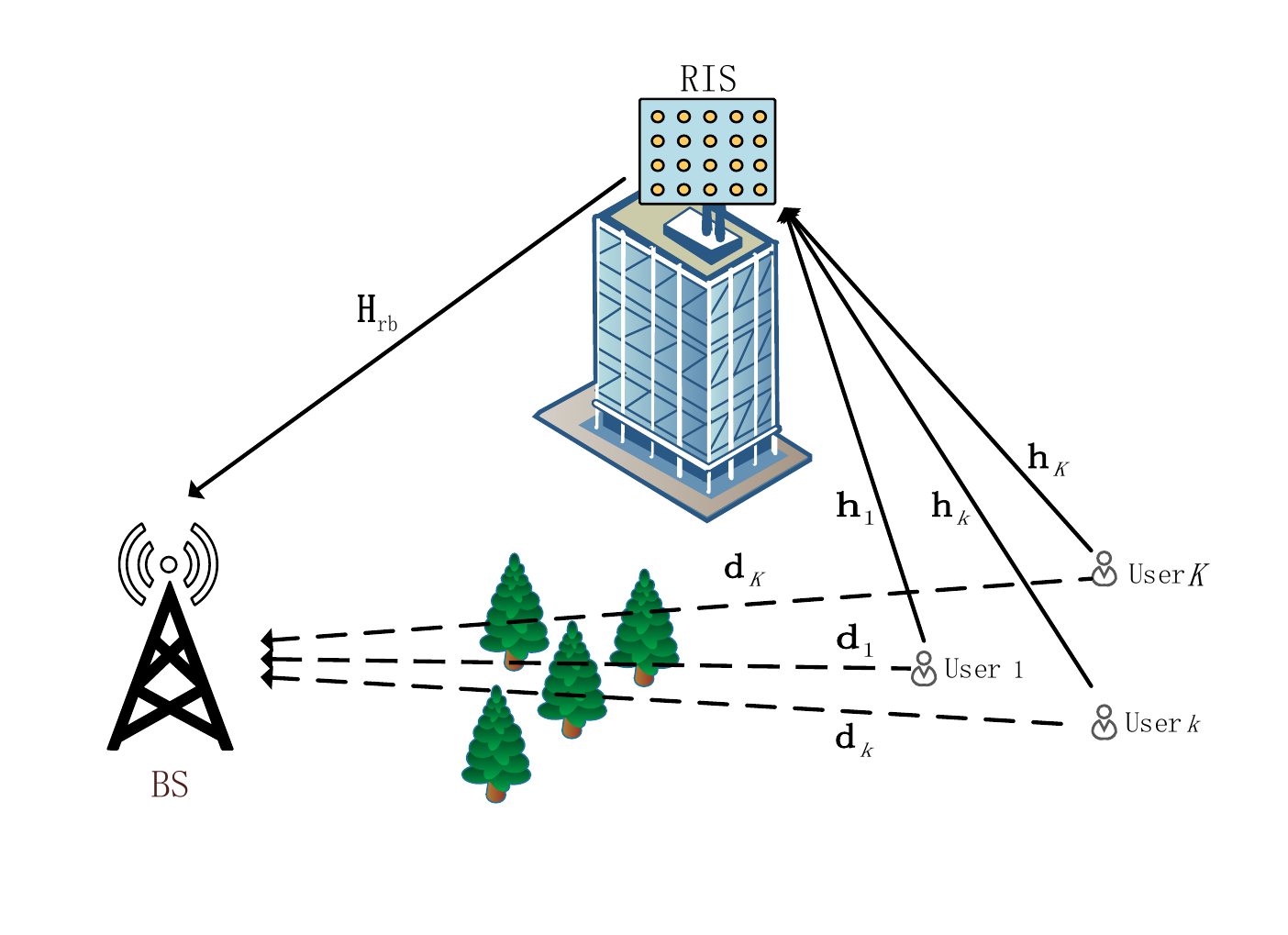}}
		 \vspace{-0.5cm}
		\caption{A typical uplink RIS-aided  MIMO system with direct links}\label{f123}
	\end{figure}
		 \vspace{-0.2cm}
	As depicted in Fig. \ref{f123}, a typical uplink RIS-aided multi-user (MU) massive MIMO communication system with direct links is considered, where a BS is equipped with $M$
	antennas and the RIS is composed of $N$ reflecting elements. The uniform square planar array (USPA) is adopted at both the BS and the RIS. $K$ single-antenna
	users communicate with BS via the RIS, and the phase shift matrix of RIS is given by
	\begin{equation}
		\bm{\Theta}=\text{diag}
		\begin{Bmatrix}
			{e^{j\theta_1},\dots,e^{j\theta_n},\dots,e^{j\theta_N}}
		\end{Bmatrix},
	\end{equation}
where  $\theta_n \in [0,2\pi)$  represents the $n$-th reflecting element of RIS.

\textcolor[rgb]{0,0,0}{Considering the imperfection of RIS, we assume that the phase noise at RIS can be written as $\hat{\theta}_n \in \mathcal{U}[-k_r\pi,k_r\pi]$ \cite{9295369}, where $\mathcal{U}$ denotes the uniform distribution and $k_r$ measures the severity of the residual impairments at the RIS. Therefore, the phase shift matrix of the RIS with phase
noise can be expressed as
		 \vspace{-0.15cm}
\begin{equation}
	\widetilde{\bm{\Theta}}=\text{diag}
	\begin{Bmatrix}
		{e^{j(\theta_1+\hat{\theta}_1)},\dots,e^{j(\theta_n+\hat{\theta}_n)},\dots,e^{j(\theta_N+\hat{\theta}_N)}}
	\end{Bmatrix}.
\end{equation}
	}
	We consider the Rician fading model for the RIS-related channel links. Specifically, the channel between the RIS and the BS is denoted by $\textbf{H}_{rb} \in \mathbb{C}^{M \times N}$, the channel between user $k$ and the RIS is denoted by $\textbf{h}_k \in \mathbb{C}^{N \times 1} $ ,  and the expressions of  
 $\textbf{H}_{rb}$ and $\textbf{h}_{k}$ are respectively given by
 		 \vspace{-0.1cm}
	\begin{align}
				\textbf{H}_{rb} &= \sqrt{\nu}
		\begin{pmatrix}
			\sqrt{\frac{\rho}{\rho +1}}\overline{\textbf{H}}_{rb} + \sqrt{\frac{1}{\rho +1}}\widetilde{\textbf{H}}_{rb}
		\end{pmatrix},\\
		\textbf{h}_k &= \sqrt{\mu_k}
		\begin{pmatrix}
			\sqrt{\frac{\epsilon_k}{\epsilon_k +1}}\overline{\textbf{h}}_k + \sqrt{\frac{1}{\epsilon_k +1}}\widetilde{\textbf{h}}_k 
		\end{pmatrix},
	\end{align}
where $\nu$ and $\mu_k$ denote the large-scale fading coefficients, $\rho$ and $\epsilon_k$ are Rician factors. $\widetilde{\textbf{H}}_{rb}$
and $\widetilde{\textbf{h}}_k \in \mathbb{C}^{N \times 1}$
are none-line-of-sight (NLoS) channel components whose
elements are independent and identical distribution (i.i.d.)
random variables following $\mathcal{CN} (0, 1)$. Correspondingly, $\overline{\textbf{H}}_{rb}$ and
$\overline{\textbf{h}}_k$ are deterministic LoS channel components,  which  can be respectively expressed as
 		 \vspace{-0.2cm}	
	\begin{align}
		\overline{\textbf{H}}_{rb} &= \textbf{a}_M\begin{pmatrix}\phi^a_{rb},\phi^e_{rb}\end{pmatrix}
		\textbf{a}_N^H\begin{pmatrix}\psi^a_{rb},\psi^e_{rb}\end{pmatrix},\\
		\overline{\textbf{h}}_k &= \textbf{a}_N	\begin{pmatrix}
			\psi^a_{kr},\psi^e_{kr}
		\end{pmatrix},
	\end{align}
where $  \psi^e_{kr}$ and $\psi^a_{kr} $ respectively denote the elevation and azimuth angles of
	arrival (AoA) from the $k$-th user to RIS, $  \phi^e_{rb}$ and $\phi^a_{rb} $ respectively represent the elevation and azimuth angles
	of departure (AoD) from the RIS to the BS, $\psi^e_{rb}$ and $\psi^a_{rb}$ respectively denote the elevation and azimuth AoA from  RIS to BS. In addition,  the channel between the RIS and the users can be written as $\textbf{H}_{ur} = [\textbf{h}_1,...,\textbf{h}_k ..., \textbf{h}_K] \in \mathbb{C}^{N \times K} $ and $\textbf{a}_Z(v^a,v^e)$ can be expressed as	
	\begin{equation}
		\begin{split}
			\textbf{a}_Z(v^a,v^e)=\big[1,...,e^{j\frac{2\pi d}{\lambda}(m\sin v^a\sin v^e + n\cos v^e)},\\
			...,e^{j\frac{2\pi d}{\lambda}\big(\big(\sqrt{Z}-1\big)\sin v^a\sin v^e + \big(\sqrt{Z}-1\big)\cos v^e\big)}\big]^T,
		\end{split}
	\end{equation}
where  $\lambda$ and  $d$ are respectively the signal wavelength and element spacing, $0\le m, n\le \sqrt{Z}-1$ \cite{2020Power}.
	
	In the direct links between the BS and users, we consider Rayleigh fading model, since rich scatters often exist on the ground. The channel of direct links is denoted by $\textbf{D} \in \mathbb{C}^{M \times K}$ that is given by
	 		 \vspace{-0.2cm}	
	\begin{equation}
		\textbf{D} = [\textbf{d}_1, ..., \textbf{d}_K],\textbf{d}_k=\sqrt{\xi_k}  \widetilde{\textbf{d}}_k ,
	\end{equation}	
where $ \xi_k $ is large-scale fading coefficient and $ \widetilde{\textbf{d}}_k$ represents the NLoS direct link for user $k$, whose elements are independent and follow the distributions of $ \mathcal{CN} (0, 1)$.

\textcolor[rgb]{0,0,0}{Therefore, the whole channel can be denoted by $\textbf{G} =\textbf{D}+\textbf{H}_{rb}\widetilde{\bm{\Theta}}\textbf{H}_{ur}$. Besides, $\textbf{g}_k \overset{\Delta}{=} \textbf{d}_k+\textbf{H}_{rb}\widetilde{\bm{\Theta}}\textbf{h}_k$
is the $k$-th column of matrix G.}

  \textcolor[rgb]{0,0,0}{The transmit distortion is denoted by $\textbf{z}_t$, whose elements are independent and follow $ \mathcal{CN} (0, k_u p_k)$, where $k_u$  measures the severity of the residual impairments at the transmitter.   The receive distortion is denoted by $\textbf{z}_r$, and follows  $ \mathcal{CN} \big(0,k_b\sum_{i=1}^{K}\widetilde{diag}\{\widetilde{\textbf{y}}_i\widetilde{\textbf{y}}_i^H\}\big)$, where $k_b $  measures the severity of the residual impairments at the receiver,  $\widetilde{\textbf{y}}_i=\textbf{g}_i(\sqrt{p_i}x_i + z_{t,i})$ and $z_{t,i}$ is the $i$-th element of $\textbf{z}_{t}$.} Here, \textcolor[rgb]{0,0,0} {$\textbf{z}_t$ models the joint effects of the non-linearities in power amplifier and digital-to-analog converters, the power amplifier noise and oscillator phase noise. $\textbf{z}_r$ models the joint effects of the non-linearities in  analog-to-digital converters and automatic gain control (AGC), the  AGC noise and the oscillator phase noise \cite{9239335}.}
	
	Based on the above definitions, the received signal can be expressed as
	 		 \vspace{-0.2cm}	
\begin{align}
			\textbf{y} &=\textbf{G}(\textbf{P}\textbf{x}+\textbf{z}_t)+\textbf{z}_r+\textbf{n} \notag \\
			&=\sum_{k=1}^{K}\textbf{g}_k (\sqrt{p_k}x_k + z_{t,k})+\textbf{z}_r+\textbf{n}.\label{e7}
		\end{align}	
	where $\textbf{P}\hspace{-0.5mm}=\hspace{-0.5mm}\text{diag}(\sqrt{p_1}, ..., \sqrt{p_K}) $  represents transmit power of
	the corresponding users, $ \textbf{x}\hspace{-1mm}=\hspace{-1mm}[x_1, ..., x_K]^T$ represents the signal vector of users, and $\mathbb{E}\{|x_k|^2\}=1$. $\textbf{n} \in \mathcal{CN} (0, \sigma^2\bm{I}_N) $ is the \textcolor[rgb]{0,0,0}{Additive white Gaussian noise.}

	\vspace{-0.1cm}
	\section{Analysis Of Uplink Achievable Rate} 
	
	To reduce the computational and implementation complexity, the low-complexity  maximal ratio combining (MRC) technique with receiver matrix $\textbf{G}^H$ is employed. Therefore, the received signal of the $k$-th user at the BS can be expressed as
	\vspace{-0.1cm}
	\begin{small}
		\begin{equation}\label{9}
			\begin{split}
				r_k&=\textbf{g}^H_k \textbf{y}\\
				&= \textbf{g}^H_k \begin{pmatrix}
					\sum_{i=1}^{K}\textbf{g}_i (\sqrt{p_i}x_i + z_{t,i})+\textbf{z}_r+\textbf{n}
				\end{pmatrix}\\
				&=\underbrace{\textbf{g}^H_k \textbf{g}_k \sqrt{p_k} x_k}_{\text{Signal}}
				+ \hspace{-0.25cm}\underbrace{\sum_{i=1,i\neq k}^{K}\hspace{-0.25cm}\textbf{g}^H_k \textbf{g}_i \sqrt{p_i} x_i}_{\text{Interference}}
				+\underbrace{\sum_{i=1}^{K}\textbf{g}^H_k \textbf{g}_i  z_{t,i} +\textbf{g}^H_k \textbf{z}_r}_{\text{HWI}}
				+\underbrace{\textbf{g}^H_k \textbf{n}}_{\text{Noise}}.
			\end{split}
		\end{equation}
	\end{small}

	By using \cite[Lemma1]{6816003}, the uplink ergodic rate of user $k$ can be approximated as
	\vspace{-0.1cm}
		\begin{equation} \label{s123}
		\begin{split}
			R_k\approx\log_2 \begin{pmatrix}
				1+\frac{p_k\mathbb{E}_{\text{signal}}^k(\bm{\Theta})}{\sum_{i=1,i\neq k}^{K}p_i\mathbb{E}_{\text{interf}}^k(\bm{\Theta}) + \mathbb{E}_{\text{hwi}}^k(\bm{\Theta})+
					\sigma^2 \mathbb{E}_{\text{noise}}^k(\bm{\Theta})}
			\end{pmatrix},
		\end{split}
	\end{equation}	
where $\mathbb{E}^k_{\text{signal}}(\bm{\Theta})$ = $\mathbb{E}\big\{||\textbf{g}_k||^4\big\}$, $\mathbb{E}^k_{\text{noise}}(\bm{\Theta})$ = $\mathbb{E}\big\{||\textbf{g}_k||^2\big\}$, $\mathbb{E}^k_{\text{interf}}(\bm{\Theta}) = \mathbb{E}\big\{|\textbf{g}_k^H\textbf{g}_i|^2\big\} $. $\mathbb{E}_{\text{signal}}^k(\bm{\Theta})$, $\mathbb{E}_{\text{interf}}^k(\bm{\Theta})$, $\mathbb{E}_{\text{noise}}^k(\bm{\Theta})$ and $\mathbb{E}_{\text{hwi}}^k(\bm{\Theta}) $ are  respectivly given by (\ref{s2123}), (\ref{s3}), (\ref{s4}) and (\ref{s5}). The derivations of the first three terms can be found in \cite{9355404} and the fourth term is proved in Appendix A. Besides, $a_k\overset{\Delta}{=}\frac{\nu\mu_k}{(\rho+1)(\epsilon_k+1)}$,  \textcolor[rgb]{0,0,0}{$\text{g}_{k_m} $is the $m$-th element of $\textbf{g}_k$, $f_k(\bm{X})\overset{\Delta}{=}\textbf{a}_N^H(\phi^a_{rb},\phi^e_{rb})\bm{X}\overline{\textbf{h}}_k$.}
 
	\newcounter{mytempeqncnt1}
	\begin{figure*}[!t]	
		\normalsize	
		\setcounter{mytempeqncnt1}{\value{equation}}
		\setcounter{equation}{\value{mytempeqncnt1}}
		\textcolor[rgb]{0,0,0}{
			\begin{equation} \label{s2123}
				\begin{split}
					\mathbb{E}^k_{\text{signal}}(\bm{\Theta})&=M^2 a^2_k \rho^2 \epsilon^2_kc_k^2+2a_kM\rho \epsilon_kc_k(\xi_k(M+1)+a_k(2M+2MN\rho+MN+MN\epsilon_k+N+N\epsilon_k+2))\\
					&
					+a^2_kM^2N^2(2\rho^2+\epsilon_k^2+2\rho\epsilon_k+2\rho+2\epsilon_k+1)+a_k^2MN^2(\epsilon_k^2+2\rho\epsilon_k+2\rho+2\epsilon_k+1)\\
					&+a_kMN(M+1)(a_k(2\rho+2\epsilon_k+1)+2\xi_k(\rho+\epsilon_k+1)+\xi_k^2(M^2+M)),
				\end{split}
			\end{equation}
			\vspace{-0.15cm}
			\begin{equation} \label{s3}
				\begin{split}
					\mathbb{E}^k_{\text{interf}}(\bm{\Theta})&=	M^2a_ka_i\rho^2\epsilon_k\epsilon_ic_kc_i+Ma_k\rho\epsilon_kc_k(a_i(\rho MN+N\epsilon_i+N+2M)+\xi_i)\\
					&+Ma_i\rho\epsilon_ic_i(a_k(\rho MN+N\epsilon_k+N+2M)+\xi_k)+MN^2a_ka_i(M\rho^2+\rho(\epsilon_i+\epsilon_k+2)\\
					&+(\epsilon_k+1)(\epsilon_i+1))
					+M^2Na_ka_i(2\rho+\epsilon_k+\epsilon_i+1)+2M^2a_ka_i\rho\epsilon_i\epsilon_k(\text{sinc}^2(k_r\pi)\text{Re}\{f^H_k(\bm{\Theta})f_i(\bm{\Theta})\overline{\textbf{h}}^H_k\overline{\textbf{h}}_i\}\\
					&+(1-\text{sinc}^2(k_r\pi))N)+M\big(a_i\xi_kN(\rho+\epsilon_i+1)+a_k\xi_iN(\rho+\epsilon_k+1)+\xi_i\xi_k\big),
				\end{split}
			\end{equation}
			\vspace{-0.3cm}	
			\begin{equation} \label{s4}
				\begin{split}
					\mathbb{E}^k_{\text{noise}}(\bm{\Theta})=M(a_k\rho\epsilon_kc_k+a_kN(\rho+\epsilon_k+1)+\xi_k),
				\end{split}
			\end{equation}
		\vspace{-0.2cm}
		where $c_k\overset{\Delta}{=}\big(\text{sinc}^2(k_r\pi)|f_k(\bm{\Theta})|^2+(1-\text{sinc}^2(k_r\pi))N\big)$, $c_i\overset{\Delta}{=}\big(\text{sinc}^2(k_r\pi)|f_i(\bm{\Theta})|^2+(1-\text{sinc}^2(k_r\pi))N\big)$.}
	
		\hrulefill
			\vspace{-0.1cm}	
			\begin{equation} \label{s5}
				\begin{split}
					\mathbb{E}^k_{\text{hwi}}(\bm{\Theta})=k_u\begin{pmatrix}
						\sum_{i=1,i\neq k}^{K} p_i \mathbb{E}^k_{\text{interf}}(\bm{\Theta}) + p_k \mathbb{E}^k_{\text{signal}}(\bm{\Theta})
					\end{pmatrix}+(1+k_u)k_b\mathbb{E}\begin{Bmatrix}
						\sum_{i=1}^{K}p_i\sum_{m=1}^{M}|\text{g}_{i_m}|^2|\text{g}_{k_m}|^2 
					\end{Bmatrix},
				\end{split}
			\end{equation}
		\vspace{-0.1cm}
			where	 $|\text{g}_{i_m}|^2|\text{g}_{k_m}|^2 $ can be expressed as			
			\textcolor[rgb]{0,0,0}{\begin{equation} \label{s6}
				\begin{split}
					|\text{g}_{i_m}|^2|\text{g}_{k_m}|^2&=a_k^2\rho^2\epsilon_k^2c_k^2\hspace{-0.2mm}+\hspace{-0.2mm}2a_k^2N(\rho^2N\hspace{-0.2mm}+\hspace{-0.2mm}\epsilon_k^2N\hspace{-0.2mm}+\hspace{-0.2mm}N\hspace{-0.2mm}+\hspace{-0.2mm}1)\hspace{-0.2mm}+\hspace{-0.2mm}4\rho a_k^2\epsilon_kN(\rho\hspace{-0.2mm}+\hspace{-0.2mm}\epsilon_k\hspace{-0.2mm}+\hspace{-0.2mm}1)c_k\hspace{-0.2mm}+\hspace{-0.2mm}4\rho a_k^2\epsilon_kN^2\hspace{-0.2mm}+\hspace{-0.2mm}a_ka_iN
					\\
					&+
					a_k^2(\rho\hspace{-0.2mm}+\hspace{-0.2mm}\epsilon_k)\hspace{-0.2mm}(2N(N+1))+\hspace{-0.2mm}2\xi_k^2\hspace{-0.2mm}+\hspace{-0.2mm}4\xi_k
					a_k(\rho\epsilon_kc_k\hspace{-0.2mm}+\hspace{-0.2mm}\rho N\hspace{-0.2mm}+\hspace{-0.2mm}\epsilon_k N\hspace{-0.2mm}+\hspace{-0.2mm}N)
					\hspace{-0.2mm}+\hspace{-0.2mm}\xi_i\xi_k\hspace{-0.2mm}+\hspace{-0.2mm}a_k\xi_i(
					\rho\epsilon_kc_k\hspace{-0.2mm}+\hspace{-0.2mm}\rho N\\
					&+\epsilon_kN\hspace{-0.2mm}+\hspace{-0.2mm}N
					)\hspace{-0.2mm}+\hspace{-0.2mm}a_ka_i\rho N^2
					\hspace{-0.2mm}+\hspace{-0.2mm}a_i\xi_k(\rho\epsilon_ic_i\hspace{-0.2mm}+\hspace{-0.2mm}\rho N\hspace{-0.2mm}+\hspace{-0.2mm}\epsilon_iN\hspace{-0.2mm}+\hspace{-0.2mm}N)\hspace{-0.2mm}+\hspace{-0.2mm}a_ka_i(\epsilon_k\hspace{-0.2mm}+\hspace{-0.2mm}1)N\big((\rho\hspace{-0.2mm}+\hspace{-0.2mm}1) N\hspace{-0.2mm}+\hspace{-0.2mm}\epsilon_i (N\hspace{-0.2mm}+\hspace{-0.2mm}1)\big)\\
					&+a_ka_i
					\rho\big(N\hspace{-0.2mm}+\hspace{-0.2mm}\epsilon_kc_k\big)\big(\rho\epsilon_ic_i\hspace{-0.2mm}+\hspace{-0.2mm}\rho N\hspace{-0.2mm}+\hspace{-0.2mm}\epsilon_i N\big)\hspace{-0.2mm}\hspace{-0.2mm}+\hspace{-0.2mm}2a_k^2(\epsilon_k\hspace{-0.2mm}+\hspace{-0.2mm}\rho)(N^2\hspace{-0.2mm}+\hspace{-0.2mm}N)+\hspace{-0.2mm}a_ka_i\rho N\big((\epsilon_k^2\hspace{-0.2mm}+\hspace{-0.2mm}\epsilon_i) c_i\hspace{-0.2mm}+\hspace{-0.2mm}2\big)\\
					&+\hspace{-0.2mm}a_ka_i\rho\epsilon_kc_k+a_ka_i\big(2\rho \epsilon_i\epsilon_k(\text{sinc}^2(k_r\pi)\text{Re}\big\{
					f^H_k(\bm{\Theta})f_i(\bm{\Theta})\overline{\textbf{h}}^H_k\overline{\textbf{h}}_i\big\}+(1-\text{sinc}^2(k_r))N) \hspace{-0.2mm}+\hspace{-0.2mm}4\rho\epsilon_kc_k\big)\hspace{-0.2mm}.	
				\end{split}
			\end{equation}
		}
		\setcounter{equation}{\value{mytempeqncnt1}}
		\hrulefill
		\vspace{-0.2cm}
	\end{figure*}
	
	\addtocounter{equation}{5}
	
	\textcolor[rgb]{0,0,0}{We assume that the transmit power is scaled with the number of antennas according to $p_k = p/M, M\to\infty,\forall k$, where $p$ is is a fixed value. For simplicity, we set $\rho=\epsilon_k=0$, i.e., only NLoS paths exist in the environment, then we have
	\begin{equation}\label{66}
		R_k \to \log_2\bigg(1+\frac{pA_1}{p(1+k_u)\sum_{i=1,i\neq k}^{K}A_2+pk_uA_1+A_3\sigma^2}\bigg),
	\end{equation}
	where $A_1=\nu\mu_kN\big(\nu\mu_kN+\nu\mu_k+2\xi_k\big)+\xi_k^2$,  $A_2=\nu^2\mu_k\mu_iN$, $A_3=\nu\mu_kN+\xi_k$.}
	
	\textcolor[rgb]{0,0,0}{From equation (\ref{66}), we can find that users in RIS-aided systems with imperfect hardware can scale down their transmit power by a factor of $1/M$ while the data rate will converge to a non-zero value as $M\to\infty$.}
	\vspace{-0.1cm}
\section{Phase Shift Optimization}
In this section, \textcolor[rgb]{0,0,0}{we aim to optimize the phase shifts of RIS to maximize the sum rate and minimum user rate}. Mathematically, the optimization problems can be formulated as follows
\vspace{-0.3cm}
	\begin{align}
		\label{16}&\max_{\bm{\Theta}}\quad \sum_{k=1}^{K}R_k,\\
		& \begin{array}{r@{\quad}r@{}l@{\quad}l}
			s.t.&\theta_n \in [0,2\pi),\forall n\\
		\end{array}.
\\\text{and}\quad&\notag\\
	\label{17}&\textcolor[rgb]{0,0,0}{\max_{\bm{\Theta}}\quad\min_{k}\quad R_k,}\\
		& \begin{array}{r@{\quad}r@{}l@{\quad}l}
			\textcolor[rgb]{0,0,0}{s.t.}&\textcolor[rgb]{0,0,0}{\theta_n \in [0,2\pi),\forall n}\\
		\end{array} .
	\end{align}
where $R_k$ is given in (\ref{s123}).

Due to the complicated data rate expression, conventional optimization techniques are not applicable. \textcolor[rgb]{0,0,0}{To address this problem, we adopt GA. We need to discretize the angle, take the phase as the chromosome and design the objective function as the fitness function.} The detailed steps of which  are given in Algorithm 1. Specifically, \textcolor[rgb]{0,0,0}{we evaluate the fitness of individuals in each generation.} Those with high fitness are retained as elites to the next generation, those with low fitness experience mutation operation to generate offspring, and those with medium fitness are used to generate parents, and then cross parents to generate offspring.  \textcolor[rgb]{0,0,0}{The complexity of the
algorithm is proportional to $qSMN^2$, where $q$ is the number
of  iterations, $S$ is the population size, $M$ is the number of BS antennas, $N$ is the number of reflecting elements \cite{8269405}.}

	\begin{algorithm}
	\caption{GA}
	\begin{algorithmic}[1]    
		\STATE Initialization: generate a population of $S=S_e+S_m+S_p$ individuals and the $i$-th individual  has a randomly generated chromosome $\bm{\Theta}_i$; the iteration
		number $q$ = 1;
		\WHILE{$q \le N*100$}
		\STATE Fitness evaluation: \textcolor[rgb]{0,0,0}{Calculate the fitness of each individual by using the objective function in (\ref{16}) or (\ref{17}) and sort them in a descending order;}
		\STATE  \textcolor[rgb]{0,0,0}{Selection: Based on the descending order, select the top $S_e$ individuals as elites;}
		\STATE  \textcolor[rgb]{0,0,0}{Mutation:  Create $S_m$ offspring from the last $S_m$ individuals} by using uniform mutation \cite{2020Power};
		\STATE Crossover: Use stochastic universal sampling \cite{2020Power} to generate $2S_p$ parents from the remaining $S_p$ individuals. Then  use two-points crossover \cite{2020Power} to create $S_p$ offspring from $2S_p$ parents;
		\STATE  Combine $S_e$ elites, $S_m+S_p$ offspring to form the next generation population; $q=q+1$;
		\ENDWHILE
		\STATE  Output the chromosome of the most fit individual in the current population.
	\end{algorithmic}
\end{algorithm}

	\section{Simulations Results} 
	\begin{figure*}[h]
		\begin{minipage}[t]{0.35\linewidth}  
			\centering
			\includegraphics[scale=0.35]{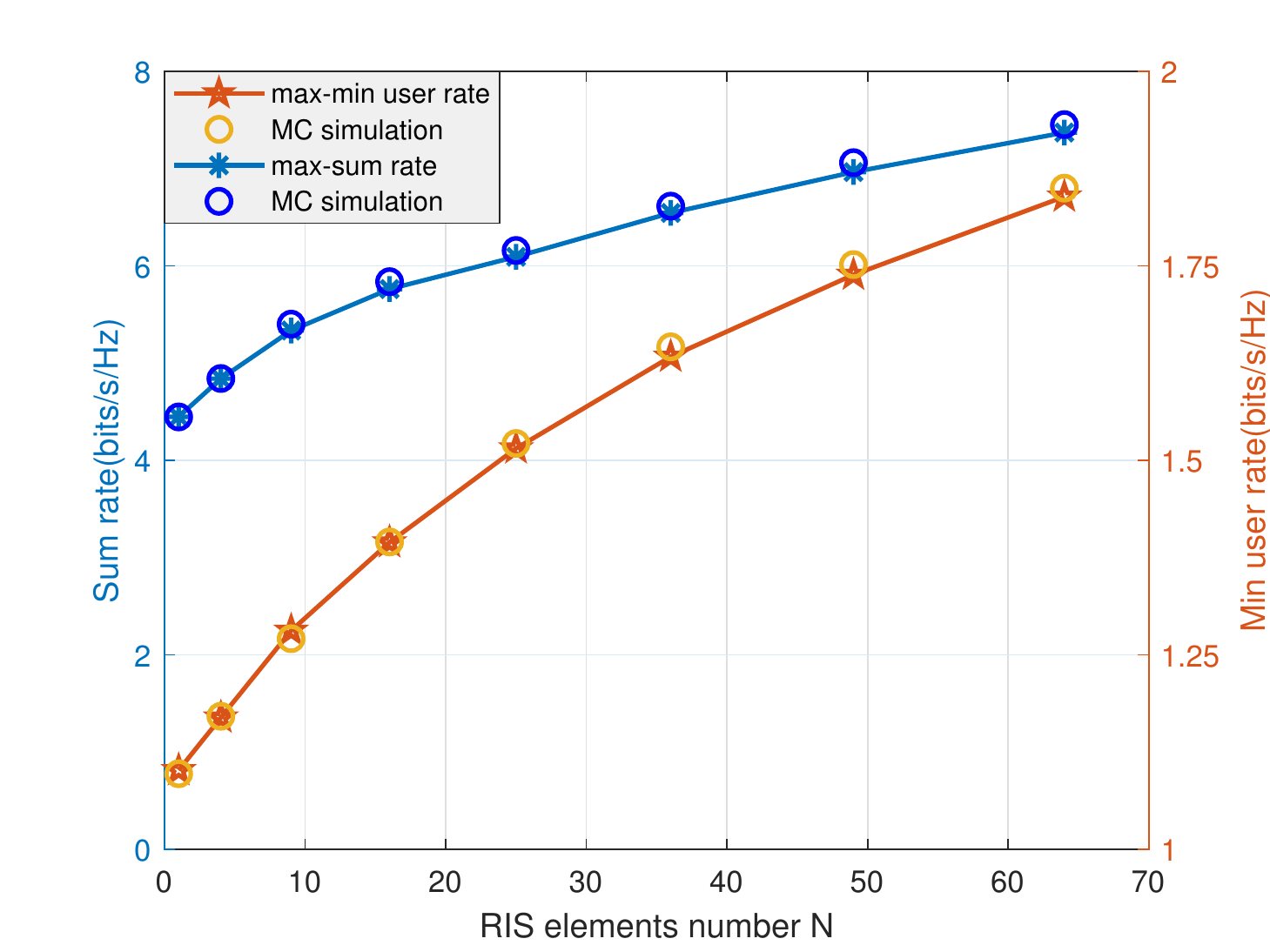} 
			\vspace{-0.3cm}
		\caption{ \hspace{0.05cm}Rate versus  $N$\quad\quad\quad.}\label{f3}
		\end{minipage}%
	\hspace{0.1cm}
		\begin{minipage}[t]{0.3\linewidth}
			\centering
			\includegraphics[scale=0.35]{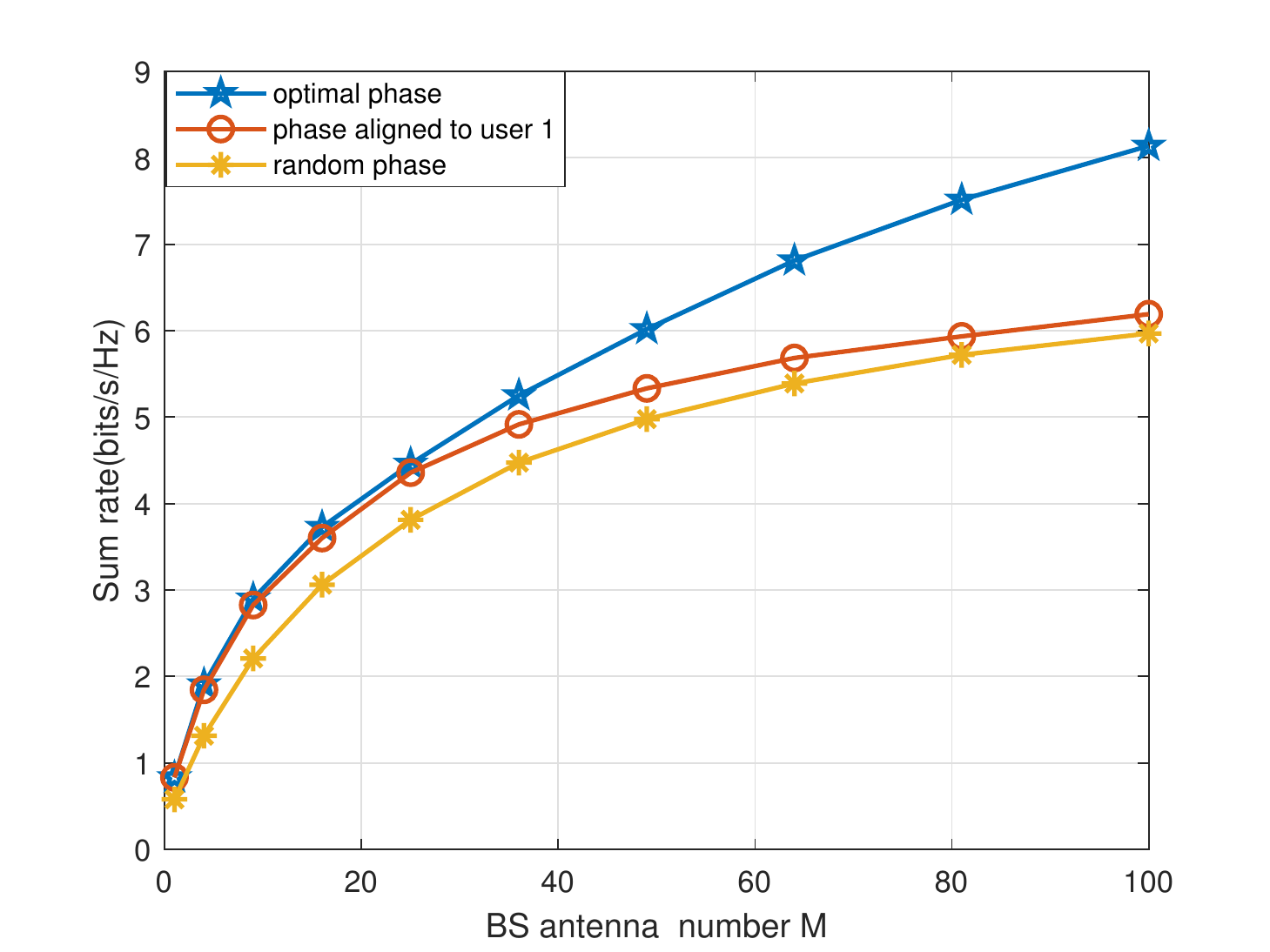}
			\vspace{-0.3cm}
			\caption{\hspace{0.05cm}Rate versus  $M$.}\label{f1}
		\end{minipage}
	\hspace{0.3cm}
		\begin{minipage}[t]{0.3\linewidth}
			\centering
			\includegraphics[scale=0.35]{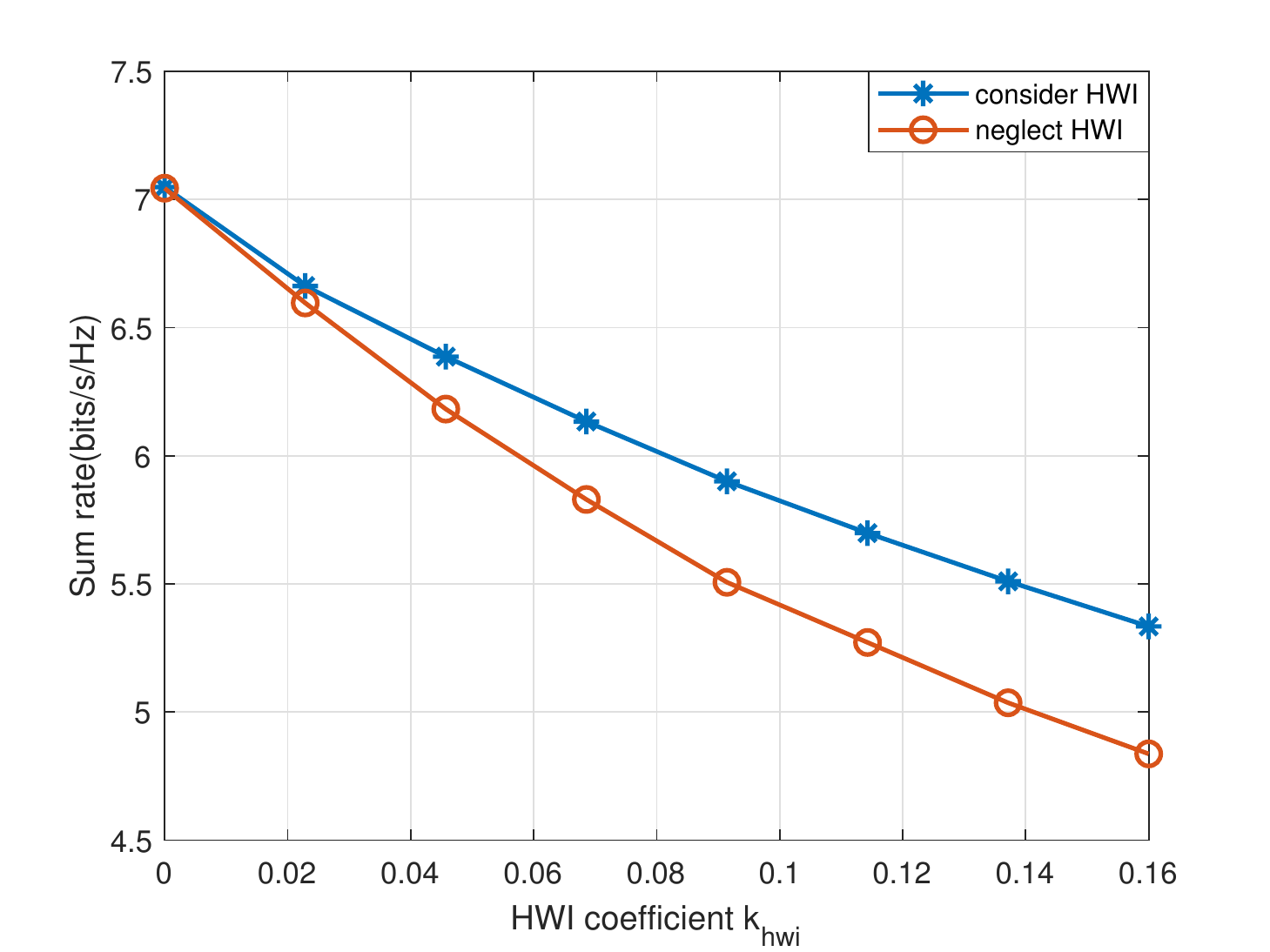}
				\vspace{-0.3cm}
			\caption{Rate versus the HWI coefficient.}\label{f2}
		\end{minipage}
	\vspace{-0.4cm}
	\end{figure*}	
In this section, extensive simulation results are provided to validate the accuracy of our analytical expression. Unless otherwise stated, \textcolor[rgb]{0,0,0} {the simulation parameters are set as follows \cite{124578}: the number of antennas of $M=50$, the number of reflecting
	elements of $N=25$, HWI coefficients $k_{r}=k_{u}=k_{b}=0.08$,} $\sigma^2=-104$ dBm, $p_k=30$ dBm, $\epsilon_k=1, \forall k, \rho=10, K=4$, RIS-BS
	distance is denoted by $l_{rb}=1000$ m. We assume that users are distributed on a semicircle with RIS as the center and a radius of 20 m. Therefore, user-RIS
	distance is denoted by $l_{ur}=20$ m, the distance between user $k$ and BS is denoted by $l_{ub}^k$ and $l_{ub}^1=l_{ub}^4=988$ m, $l_{ub}^2=l_{ub}^3=980$ m. The large-scale fading coefficients are $\mu_k=10^{-3}l_{ur}^{-2}, \nu=10^{-3}l_{rb}^{-2.5}, \xi_k=10^{-3}(l_{ub}^k)^{-4}, \forall k$ . The AoA and the AoD are all random values within $[0,2\pi]$. 
	
	\textcolor[rgb]{0,0,0}{Fig. \ref{f3} depicts the max-sum rate and max-minimum user rate.} The simulation shows that the Monte-Carlo (MC) simulation results are consistent with the derived results, which verify the correctness of the derived expression.

%

	\textcolor[rgb]{0,0,0}{In Fig. \ref{f1}, we compare the system performance in different scenarios. Obviously, with the increase of $M$, the performance gain of optimal phase shifts will become more and more prominent, which demonstrates  the superiority of using GA to optimize the phase shifts in massive MIMO systems}.

	\textcolor[rgb]{0,0,0}{Fig. \ref{f2} shows the significance of investigating HWIs. In Fig. \ref{f2}, one of the curves considers  HWIs, while the other neglects HWIs to optimize the phase shift and substitute the obtained phase shift solution into the actual system with HWIs. $k_{r},k_{u},k_{b}=k_{hwi}$. The simulation shows that with the increase of HWI coefficients, the performance gap between the two schemes becomes larger}.
%
	\section{Conclusion}
	We investigated an RIS-aided MU massive MIMO system with HWIs. The approximate analytical expression of uplink achievable rate has been derived based on Rician fading channel and MRC technique. Through MC simulation, we verified the accuracy of analytical expression. \textcolor[rgb]{0,0,0}{In addition, we showed that it is crucial to take HWIs into account  when designing the phase shift of RIS. We will consider the spatial correlation at the RIS and the transmit power allocation in our future work.}

	\appendices 	
	\section{Derivations Of $\mathbb{E}^k_{\text{hwi}}(\bm{\Theta})$} 		
	\vspace{-0.2cm}
	\begin{equation}
		\begin{split}
		&\mathbb{E}\big\{|\textbf{g}_k^H\textbf{z}_r|^2\}\\	
		\overset{\alpha}{=}&\enspace k_b(1+k_u)
		\sum_{i=1}^{K}p_i\sum_{m=1}^{M}\mathbb{E}\begin{Bmatrix}|\text{g}_{i_m}|^2|\text{g}_{k_m}|^2 
		\end{Bmatrix},
	\end{split}
	\end{equation}
where ($\alpha$) is obtained by removing the zero terms.	
	
	Then, we have
	\begin{equation}
		\begin{split}
			&\enspace\sum_{i=1}^{K}p_i\sum_{m=1}^{M}\mathbb{E}\begin{Bmatrix}|\text{g}_{i_m}|^2|\text{g}_{k_m}|^2 
			\end{Bmatrix}\\
			=&\enspace
			p_k\hspace{-0.1cm}\sum_{m=1}^{M}\mathbb{E}\begin{Bmatrix}
				|\text{g}_{k_m}|^4
			\end{Bmatrix}\hspace{-0.1cm}+\hspace{-0.3cm}\sum_{i=1,i\neq k}^{K}\hspace{-0.1cm}p_i\hspace{-0.1cm}\sum_{m=1}^{M}\mathbb{E}\begin{Bmatrix}
				|\text{g}_{i_m}|^2|\text{g}_{k_m}|^2
			\end{Bmatrix}.
		\end{split}
	\end{equation}

\textcolor[rgb]{0,0,0}{Let $\textbf{v}_k=\textbf{H}_{rb}\widetilde{\bm{\Theta}}\textbf{h}_k$, we can rewrite $\text{v}_{k_m}$ in the  form of (\ref{s9}) at the top of the next page, where  $\text{ v}_{ k_m}$ and $\text{ a}_{ M_m}(\psi^a_{rb},\psi^e_{rb})$ are the $m$-th element of respective vectors, $\big[\widetilde{\textbf{H}}_{rb}\big]_{mn}$ denotes the $(m,n)$-th entry of matrix $\widetilde{\textbf{H}}_{rb}$, $\widetilde{\textbf{h}}_{k_m} $represents the $m$-th element of $\widetilde{\textbf{h}}_{k}$.}
		\vspace{-0.3cm}
	\subsection{Derivations of \enspace $\mathbb{E}\big\{|\text{g}_{k_m}|^4\big\} $}
	\newcounter{mytempeqncnt2}
	\begin{figure*}[!t]	
		\normalsize	
		\setcounter{mytempeqncnt2}{\value{equation}}
		\setcounter{equation}{\value{mytempeqncnt2}}
		\begin{equation}\label{s9}
			\begin{split}
				\text{ v}_{ k_m}&=\sqrt{\frac{\nu\mu_k}{(\rho+1)(\epsilon_k+1)}}\Big(\underbrace{\sqrt{\rho\epsilon_k}\text{ a}_{Mm}\begin{pmatrix}\psi^a_{rb},\psi^e_{rb}\end{pmatrix}\textcolor[rgb]{0,0,0}{f_k(\bm{\widetilde{\Theta}})}}_{\text{ v}_{k_m}^1}
				+\underbrace{\sqrt{\rho}\text{ a}_{Mm}\begin{pmatrix}\psi^a_{rb},\psi^e_{rb}\end{pmatrix}\textstyle\sum_{n=1}^{N}\hspace{-0.1cm}\text{ a}_{Nn}^*\begin{pmatrix}\phi^a_{rb},\phi^e_{rb}\end{pmatrix}e^{j\textcolor[rgb]{0,0,0}{(\theta_n+\hat{\theta}_n)}}\widetilde{h}_{k_n}}_{\text {v}_{k_m}^2}\\
				&+\underbrace{\sqrt{\epsilon_k}\textstyle\sum_{n=1}^{N}\big[\widetilde{\textbf{H}}_{rb}\big]_{mn}e^{j\textcolor[rgb]{0,0,0}{(\theta_n+\hat{\theta}_n)}}\text{ a}_N	\begin{pmatrix}
						\psi^a_{kr},\psi^e_{kr}
				\end{pmatrix}}_{\text{ v}_{k_m}^3}
				+\underbrace{\textstyle\sum_{n=1}^{N}\big[\widetilde{\textbf{H}}_{rb}\big]_{mn}e^{j\textcolor[rgb]{0,0,0}{(\theta_n+\hat{\theta}_n)}}\widetilde{h}_{k_n}}_{\text{ v}_{k_m}^4}
				\Big)
			\end{split}
		\end{equation}
		\setcounter{equation}{\value{mytempeqncnt2}}
		\vspace{-0.15cm}
		\hrulefill
		\vspace{-0.1cm}
	\end{figure*}
	
	\addtocounter{equation}{1}
	
	We have	
	\begin{equation}\label{s00}
		\begin{split}
			&\enspace\mathbb{E}\big\{|\text{g}_{k_m}|^4\big\}\\
			= &\enspace2\xi_k^2+\mathbb{E}\big\{|\text{v}_{k_m}|^4\big\}+2\xi_k\mathbb{E}\big\{|\text{v}_{k_m}|^2\big\}\\
			 &\enspace+4\mathbb{E}\Big\{\big(\text{Re}\big\{(\text{v}_{k_m})(\text{d}_{k_m})^*\big\}\big)^2\Big\},
		\end{split}	
	\end{equation}
where \textcolor[rgb]{0,0,0}{$\mathbb{E}\big\{|\text{v}_{k_m}|^2\big\}=a_k\big(\rho\epsilon_kc_k+\rho N+\epsilon_k N+N\big)$} can be easily obtained,   $\mathbb{E}\big\{|\text{v}_{k_m}|^4\big\}$ was derived in \cite{2020Power} and $\mathbb{E}\Big\{\big(\text{Re}\big\{(\text{v}_{k_m})(\text{d}_{k_m})^*\big\}\big)^2\Big\}$ can be expressed as

	\begin{equation}\label{s0}
		\begin{split}
			&\enspace\mathbb{E}\Big\{\big(\text{Re}\big\{(\text{v}_{k_m})(\text{d}_{k_m})^*\big\}\big)^2\Big\}\\
			=&\enspace a_k\xi_k\Big(\mathbb{E}\Big\{\big(\text{Re}\big\{(\text{v}_{k_m}^1)(\widetilde{\text{d}}_{k_m})^*\big\}\big)^2\Big\}
			\hspace{-0.1cm}+\hspace{-0.1cm}\mathbb{E}\Big\{\big(\text{Re}\big\{(\text{v}_{k_m}^2)(\widetilde{\text{d}}_{k_m})^*\big\}\big)^2\Big\}\\
			&\enspace+\mathbb{E}\Big\{\big(\text{Re}\big\{(\text{v}_{k_m}^3)(\widetilde{\text{d}}_{k_m})^*\big\}\big)^2\Big\}
		\hspace{-0.1cm}+\hspace{-0.1cm}\mathbb{E}\Big\{\big(\text{Re}\big\{(\text{v}_{k_m}^4)(\widetilde{\text{d}}_{k_m})^*\big\}\big)^2\Big\}\Big).
		\end{split}
	\end{equation}

	Assume $\text{v}_{k_m}^1=s+jt, \widetilde{d}_{k_m}=p+jq$, where\enspace $p\in\mathcal{CN}(0,\frac{1}{2})$, $q \in \mathcal{CN}(0,\frac{1}{2})$. 
	Thus, we can derive
	\begin{equation}\label{s2}
		\begin{split}
			&\enspace\mathbb{E}\Big\{\big(\text{Re}\big\{(\text{v}_{k_m}^1)(\widetilde{\text{d}}_{k_m})^*\big\}\big)^2\Big\}
			= \textcolor[rgb]{0,0,0}{\enspace\frac{1}{2}\rho\epsilon_kc_k.}\\
			&\enspace\mathbb{E}\Big\{\big(\text{Re}\big\{(\text{v}_{k_m}^2)(\widetilde{\text{d}}_{k_m})^*\big\}\big)^2\Big\}
			=\frac{1}{2}\rho N.\\
			&\enspace\mathbb{E}\Big\{\big(\text{Re}\big\{(\text{v}_{k_m}^3)(\widetilde{\text{d}}_{k_m})^*\big\}\big)^2\Big\}
			=\frac{1}{2}\epsilon_k N.\\
			&\enspace\mathbb{E}\Big\{\big(\text{Re}\big\{(\text{v}_{k_m}^4)(\widetilde{\text{d}}_{k_m})^*\big\}\big)^2\Big\}
			=\frac{1}{2}N.
		\end{split}
	\end{equation}
\quad Substituting (\ref{s2})  into (\ref{s0}), we complete the derivations of $\mathbb{E}\big\{|\text{g}_{k_m}|^4\big\} $.
	
		\vspace{-0.1cm}
	\subsection{Derivations of \enspace $\mathbb{E}\big\{|\text{g}_{i_m}|^2|\text{g}_{k_m}|^2\big\}, \forall i \ne k $}
		\vspace{-0.5cm}
	\begin{equation}\label{s23}
		\begin{split}
			&\enspace\mathbb{E}\big\{
			|\text{g}_{i_m}|^2|\text{g}_{k_m}|^2\big\}\\
			\overset{\beta}{=} &\enspace\mathbb{E}\big\{|\text{d}_{i_m}|^2|\text{d}_{k_m}|^2\big\}+\mathbb{E}\big\{|\text{d}_{i_m}|^2|\text{v}_{k_m}|^2\big\}\\
			&\enspace+\mathbb{E}\big\{|\text{d}_{k_m}|^2|\text{v}_{i_m}|^2\big\}+\mathbb{E}\big\{|\text{v}_{i_m}|^2|\text{v}_{k_m}|^2\big\},
		\end{split}
	\end{equation}
where ($\beta$) is obtained by removing the zero terms.
	
	We can readily obtain
	\begin{equation}
		\begin{split}
			&\enspace\mathbb{E}\big\{|\text{d}_{i_m}|^2|\text{d}_{k_m}|^2\big\}=\xi_i\xi_k,\\
			&\enspace\mathbb{E}\big\{|\text{d}_{i_m}|^2|\text{v}_{k_m}|^2\big\}=\xi_i\mathbb{E}\big\{|\text{v}_{k_m}|^2\big\},\\
			&\enspace\mathbb{E}\big\{|\text{d}_{k_m}|^2|\text{v}_{i_m}|^2\big\}=\xi_i\mathbb{E}\big\{|\text{v}_{i_m}|^2\big\}.\\
		\end{split}
	\end{equation}

	The expression of $\mathbb{E}\big\{|\text{v}_{i_m}|^2|\text{v}_{k_m}|^2\big\}$ is given by (\ref{s25}) at the top of the this page and we will calculate the  terms in (\ref{s25}).
	
	\newcounter{mytempeqncnt3}
	\begin{figure*}[!t]	
		\normalsize	
		\setcounter{mytempeqncnt3}{\value{equation}}
		\setcounter{equation}{\value{mytempeqncnt3}}
		\vspace{-0.3cm}
		\begin{equation}\label{s25}
			\begin{split}
				&\enspace\mathbb{E}\big\{|\text{v}_{i_m}|^2|\text{v}_{k_m}|^2\big\}\\
				&\overset{\gamma}{=} \enspace 4a_ka_i\Big(\mathbb{E}\Big\{\text{Re}\big\{\text{v}_{k_m}^1(\text{v}_{k_m}^3)^*\big\}\text{Re}\big\{\text{v}_{i_m}^1(\text{v}_{i_m}^3)^*\big\}\Big\}+\mathbb{E}\Big\{\text{Re}\big\{\text{v}_{k_m}^1(\text{v}_{k_m}^3)^*\big\}\text{Re}\big\{\text{v}_{i_m}^2(\text{v}_{i_m}^4)^*\big\}\Big\}\\
				&
				+\mathbb{E}\Big\{\text{Re}\big\{\text{v}_{k_m}^2(\text{v}_{k_m}^4)^*\big\}\text{Re}\big\{\text{v}_{i_m}^1(\text{v}_{i_m}^3)^*\big\}\Big\}+\mathbb{E}\Big\{\text{Re}\big\{\text{v}_{k_m}^2(\text{v}_{k_m}^4)^*\big\}\text{Re}\big\{\text{v}_{i_m}^2(\text{v}_{i_m}^4)^*\big\}\Big\}\Big)+a_ka_i\mathbb{E}\Big\{\sum_{\omega=1}^{4}|\text{v}_{k_m}^\omega|^2\sum_{\omega=1}^{4}|\text{v}_{i_m}^\omega|^2\Big\},
			\end{split}
		\end{equation}
		where ($\gamma$) is obtained by removing the zero terms.
		
		\setcounter{equation}{\value{mytempeqncnt3}}
			\vspace{-0.2cm}
		\hrulefill
		\vspace{-0.3cm}
	\end{figure*}
	
	\addtocounter{equation}{1}
	
	The first one is
		\textcolor[rgb]{0,0,0}{\begin{equation}\label{z1}
			\begin{split}
				&\enspace\mathbb{E}\Big\{\sum_{\omega=1}^{4}|\text{v}_{k_m}^\omega|^2\sum_{\omega=1}^{4}|\text{v}_{i_m}^\omega|^2\Big\}\\
				=&\enspace\rho\epsilon_kc_k\big(\rho\epsilon_ic_i+\rho N+\epsilon_i N+1\big)+\rho N\big(\rho\epsilon_ic_i+\rho N\\
				&+\epsilon_i N+N\big)+\epsilon_k N\big(\rho\epsilon_kc_i+\rho N+\epsilon_i (N+1)\\
				&+N\big)+N\big(\rho\epsilon_ic_i+\rho N+\epsilon_i (N+1)+N+1\big).
			\end{split}
		\end{equation}}
	\vspace{-0.2cm}
	Assume that
\begin{equation}
	\begin{split}
		&\text{a}_{Mm}(\psi_{rb}^a,\psi_{rb}^e)\textcolor[rgb]{0,0,0}{f_k(\bm{\widetilde{\Theta}})e^{-j(\theta_n+\hat{\theta}_n)}}\text{a}_{Nn}^*(\psi_{kr}^a,\psi_{kr}^e)=\sigma_c^{kn}+j\sigma_s^{kn},\\
		&\text{a}_{Mm}(\psi_{rb}^a,\psi_{rb}^e)\textcolor[rgb]{0,0,0}{f_i(\bm{\widetilde{\Theta}})e^{-j(\theta_n+\hat{\theta}_n)}}\text{a}_{Nn}^*(\psi_{ir}^a,\psi_{ir}^e)=\sigma_c^{in}+j\sigma_s^{in},\\
		&\big[\widetilde{\textbf{H}}_{rb}\big]_{mn}=s_{mn}+jt_{mn}.
	\end{split}
\end{equation}
\vspace{-0.2cm}
Thus, the second one is
	\begin{equation}\label{z2}
		\begin{split}
			&\enspace\mathbb{E}\Big\{\text{Re}\big\{\text{v}_{k_m}^1(\text{v}_{k_m}^3)^*\big\}\text{Re}\big\{\text{v}_{i_m}^1(\text{v}_{i_m}^3)^*\big\}\Big\}\\
			=&\enspace\rho\epsilon_i\epsilon_k\mathbb{E}\Big\{\sum_{n=1}^{N}\sigma_c^{kn}\sigma_c^{in}s_{mn}^2+\sigma_s^{kn}\sigma_s^{in}t_{mn}^2\Big\}\\
			=&\enspace\frac{\rho\epsilon_i\epsilon_k}{2}\textcolor[rgb]{0,0,0}{\big(\text{sinc}^2(k_r\pi)\text{Re}\big\{f_k(\bm{\Theta})\overline{\textbf{h}}_k^H\overline{\textbf{h}}_if_i^*(\bm{\Theta})\big\}}\\
			&\textcolor[rgb]{0,0,0}{+(1-\text{sinc}^2(k_r\pi))N\big).}
		\end{split}
	\end{equation}

	Likewise, we have
	\begin{equation}\label{z3}
		\begin{split}
			\mathbb{E}\Big\{\text{Re}\big\{\text{v}_{k_m}^1(\text{v}_{k_m}^3)^*\big\}\text{Re}\big\{\text{v}_{i_m}^2(\text{v}_{i_m}^4)^*\big\}\Big\}
			=\textcolor[rgb]{0,0,0}{\frac{\rho\epsilon_k}{2}c_k,}
		\end{split}
	\end{equation}
\vspace{-0.5cm}
	\begin{equation}\label{z4}
		\begin{split}
			\mathbb{E}\Big\{\text{Re}\big\{\text{v}_{k_m}^2(\text{v}_{k_m}^4)^*\big\}\text{Re}\big\{\text{v}_{i_m}^1(\text{v}_{i_m}^3)^*\big\}\Big\}= \textcolor[rgb]{0,0,0}{\frac{\rho\epsilon_i}{2}c_i,}
		\end{split}
	\end{equation}
\vspace{-0.5cm}
	\begin{equation}\label{z5}
		\begin{split}
			\mathbb{E}\Big\{\text{Re}\big\{\text{v}_{k_m}^2(\text{v}_{k_m}^4)^*\big\}\text{Re}\big\{\text{v}_{i_m}^2(\text{v}_{i_m}^4)^*\big\}\Big\}
			=\frac{\rho N}{2}.
		\end{split}
	\end{equation}

	Substituting (\ref{z1}), (\ref{z2}), (\ref{z3}), (\ref{z4})  and (\ref{z5}) into (\ref{s25}), we complete the derivations of $\mathbb{E}\big\{|\text{g}_{i_m}|^2|\text{g}_{k_m}|^2\big\} $.
	
	Finally, substituting $\mathbb{E}\big\{|\text{g}_{i_m}|^2|\text{g}_{k_m}|^2\big\} $ and $\mathbb{E}\big\{|\text{g}_{k_m}|^4\big\} $ into (\ref{s6}), we complete the proof of $\mathbb{E}^k_{hwi}(\bm{\Theta})$.

	


	\vspace{0.1cm}
	\bibliography{text}
\end{document}